\documentclass[onecolumn,preprintnumbers,amsmath,amssymb,floatfix,superscriptaddress]{revtex4}
\usepackage{graphicx}
\usepackage{dcolumn}
\usepackage{epstopdf}
\usepackage{bm}
\usepackage{amsmath}
\usepackage{amssymb}   

\usepackage{bm} 
\usepackage{dcolumn}
\usepackage{color}
\usepackage{mathrsfs}
\usepackage{mathtools}
\usepackage{amsfonts}
\usepackage{varioref}
\usepackage{epstopdf}
\usepackage{pdfpages}
\usepackage{float}
\usepackage[dvipsnames]{xcolor}
\usepackage{color,soul}
\usepackage{makeidx}
\usepackage{showlabels}
\graphicspath{ {/home/} }
\RequirePackage[colorlinks,citecolor=blue,urlcolor=magenta,linkcolor=blue]{hyperref}

\begin{document}

	\title{\large \bf Remixing of a phase separated binary colloidal system with particles of different sizes in an external modulation}

	\author{\bf Suravi Pal}
	\affiliation{Physics of complex systems, S. N. Bose National Centre for Basic Sciences, JD Block, Sector-III, Salt Lake, Kolkata 700106, India \\{suravipal@bose.res.in}\\{\bf \large and}}
	
	\author{\bf J. Chakrabarti}
	\affiliation{Physics of complex systems, S. N. Bose National Centre for Basic Sciences, JD Block, Sector-III, Salt Lake, Kolkata 700106, India \\{jaydeb@bose.res.in}}

	\author{\bf Srabani Chakrabarty nee Sarkar}
	\affiliation{Department of Physics, Lady Brabourne College, P-1/2, Suhrawardy Ave, Beniapukur, Kolkata, West Bengal 700017, India \\{srabanichakrabarty65@gmail.com}}

	\begin{abstract}
		We explore  phase behaviour of a binary colloidal system under external spatially periodic modulation. We perform Monte Carlo simulation on a binary mixture of big and small repulsive Lennard-Jones particles with diameter ratio 1:2. We characterise structure by isotropic and anisotropic pair correlation function, cluster size distribution, bond angle distribution, order parameter and specific heat. We observe demixing of the species in the absence of the external modulation. However, mixing of the species gets enhanced with increasing potential strength. The de-mixing order parameter shows discontinuity and the specific heat shows a peak with increasing modulation strength, characterizing a first order phase transition
	\end{abstract}
	
	\maketitle 
	

\section{Introduction}

Mixing of different components in condensed phase is of paramount importance in systems ranging from alloys\cite{paper 1, paper 2, paper 3} to those as complicated as biological cells\cite{paper 4, paper 5}. Tuning miscibility among macro-molecules is thus an active area of research, spanning a diverse areas as well as technological applications. Here we examine the mixing property of a binary colloidal macro-molecular system in the presence of an external modulation potential.

Colloids  are ideal systems to explore condensed phase properties at microscopic level following the individual particle motions. This is so because the colloidal particles  can be probed via laser light scattering and optical microscopy \cite{paper 6, paper 7, paper 8} due to their length size and slow movement. Colloids can be easily modulated  by external perturbations, like laser field, electric field\cite{paper 9}, magnetic field\cite{paper 10} and  shear\cite{paper 11}. Modulated structures are often important for technological applications as well. Understanding colloidal systems under external perturbations from microscopic considerations has drawn considerable research interests\cite{paper 12, paper 12, paper 13, paper 14, paper 15, paper 16, paper 17}. Modulated colloids often show nontrivial behaviour. For instance, a mono-disperse colloidal system subject to a one-dimensional stationary laser modulation with wavelength matching with the mean inter-particle separation undergo re-entrant melting with increasing modulation strength \cite{paper 18, paper 19, paper 20}. Many interesting studies are reported on colloids under external potential\cite{paper 21, paper 22, paper 23}, such as  two-dimensional melting behavior of super-paramagnetic colloidal particles under quenched disorder\cite{paper 24},  freezing and melting of a colloidal adsorbate on a one dimensional quasi-crystalline substrate\cite{paper 25}, effective forces in modulated colloids\cite{paper 26}, modulated phases in dense colloids \cite{paper 27}, heterogeneous dynamics of colloids under external potential \cite{paper 28, paper 29, paper 30} and so on. 

Experiments on a binary colloidal system with particles of different sizes subject to a spatially periodic potential has also been reported  using the external modulation generated in a given direction by interfering laser beams\cite{paper 31}. The wavelength of the modulation is equal to the size of the bigger particles, and the modulation is stronger for them than the smaller particles. The arrangement of the particles are recorded using optical microscopy and characterized in terms of various static quantities. It is observed that an increase in the external modulation  amplitude leads to  localization of the large particles analogous to a modulated liquid\cite{paper 32} with an increasing fraction of small particles caged by the bigger particles. The smaller particles arrange themselves in a triangular fashion inside those cages.
The large particles are observed to remains ordered, whereas the lattice of small particles became disordered upon increasing the temperature. 

Numerous demixing studies have been reported on colloidal mixtures. There have been studies on demixing of model colloids composed of two different sizes of polystyrene spheres using diffusing wave  spectroscopy \cite{paper 33}. Experiments on a binary dispersion of like-charged colloidal particles with large charge asymmetry but similar size exhibit phase separation into crystal and fluid phases under very low salt conditions. Here the colloid–ion interactions provide a driving force for crystallization of one species\cite{paper 34}. Confocal-microscopy studies on demixing and remixing with temperature in binary liquids containing colloidal particles has been reported \cite{paper 35, paper 36}. However, there has been no study yet to the best of our knowledge how an external modulation changes the demixing behaviour in a binary colloid. 

In this backdrop we study a model binary system with different particles of  diameter ratio 1:2 in an external modulation as in Ref.\cite{paper 37}. The external modulation  wavelength is equal to the diameter of the big particles and the strength of modulation is twice for the big particles than the small ones. We study the system using the Monte Carlo (MC) simulation at room temperature where the Metropolis sampling has been performed based on the energy cost of a particle movement due to interaction with all other particles and the external modulation. For a fixed 1:1 composition and particle density as in Ref.\cite{paper 37}, we vary the external modulation strength. We calculate pair correlation function, cluster size distribution and  bond angle parameter  to characterize structural changes in the system with the strength of external modulation. We also compute thermodynamic quantities such as demixing order parameter and specific heat in order to investigate the phase transition as the structural changes occur in the system.  

We observe that the two components in the system undergo phase separation in absence of any external modulation at room temperature.  presence of external potential, the bigger particles align themselves along the potential minima due to the matching of their diameter with the wavelength of the external potential and stronger modulation strength, while the clusters of the smaller particles break into smaller clusters. We observe hexagonal ordering among the small particles within the clusters. The enhanced remixing of the species with increasing field strengths is reflected in the decrease of the mean cluster size and the shift of the mean demixing order parameter to lower values with discontinuity as a function of the external potential.  The specific heat shows peak as a function of the modulation strength, suggesting the presence of a thermodynamic first order phase transition.

\section{System details}	

The particles are taken to interact via repulsive potential of the form:
\begin{equation}
V^{(\alpha\beta)}_{ij}(r)=4\epsilon_{\alpha\beta} (\frac{\sigma_{\alpha\beta}}{r_{ij}})^{12}
\end{equation}
\noindent Here $r_{ij}$ is the distance between $i^{th}$ and $j^{th}$ particles belonging to species $\alpha(=b,s)$ and $\beta(=b,s) $ respectively.  Here $\alpha=\beta$ corresponds to the particles of the same species, while $\alpha\neq\beta$ corresponds to the cross species interactions. We take $\sigma_{bb}=D_{b},\sigma_{ss}=D_{s}$ and $\sigma_{bs}=0.5(\sigma_{bb}+\sigma_{ss})$, D$_{b}$ and D$_{s}$ being the diameter of the big and small particles respectively. We fix  $\frac{\epsilon_{bb}}{K_BT}=\frac{\epsilon_{ss}}{K_BT}=1.0$ and  vary $\frac{\epsilon_{bs}}{K_BT}$. 
We further subject the system to an external spatially periodic potential of the form:
\begin{equation}
V^{\alpha}_{ext}(x)=-V_0^{\alpha}cos(\frac{2\pi x}{\lambda}),
\end{equation}
where $x$ denote the x-coordinate of the particles, $\lambda$  the wavelength of the external potential equal to the diameter of the bigger particles, and $V_0^b= V_0= 2V_{0}^s$.

Here $D_b(=5\mu m)$ and $D_s(=2.5\mu m)$. The packing fraction, $\eta(=N_d/4\pi D_b^2+N_s/4D_s^2)/A$, where $A$ is the area of the system) = 0.72. We take $D_s$ as the unit of length and $K_BT$ for room temperature as the unit of energy. Monte Carlo (MC) simulations are carried out on a fixed N number of particles with equal number of big and small particles in a box of reduced length in x-direction, $\frac{L_x}{D_s}$ and that in y-direction $L_y=\frac{\sqrt{3}}{2}L_x$.  We run for a total of $10^6$ MC steps where the equilibration is judged from the energy values. Different quantities of interests are calculated over the equilibrated trajectories and averaged over five independent runs. Most of the results are reported on N=1024 particles. We also study the system for N=144 and N=4096  keeping the area fraction fixed to check the size dependence around the structural changes.

\section{Results and discussions}	

{\it Structure without modulation}

Let us consider  the system in absence of external potential$(\beta V_0=0.0)$. Fig. 1(a) shows a typical particle arrangement in equilibrium. We observe clusters of small particles in the background  of the big particles. We characterise the structural correlations from the histograms of the separation between particle pairs, also called the radial distribution function (rdf) $g_{\alpha\beta}(r)$\cite{paper 38}. This quantity describes the probability of finding a pair of particles belonging to species $\alpha$ and $\beta$ at separation $r$.  Fig. 1 (b) shows the rdf for big-big $g_{bb}(r)$, small-small $g_{ss}(r)$, big-small $g_{bs}(r)$ pair of particles in absence of external potential.  We observe liquid-like short ranges structure behaviour in both $g_{bb}(r)$ and $g_{ss}(r)$. However, $g_{bs}(r)$ is much weaker. This means that the two species do not find in the vicinity of each other due to phase separation between them.

{\it Modulated system}

Let us now consider the effect of the external potential. We show equilibrated snapshot for $\beta V_0=5.0$ in fig. 2(a). The snapshot shows that the big particles are aligned along the potential minima due to the matching of wavelength of external potential to the diameter of the bigger particles and stronger modulation strength than the smaller particles. The clusters of the smaller particles break into smaller clusters to make way for the bigger particles. 

The anisotropic structure of the system is characterized by the pair correlation functions (PCF) in Figs. 2(b)-(d). They  are obtained by binning the pair separations both in x- and y-directions. It may be noted that the rdf is the circular symmetric counter-part of the PCF. $g_{bb}(x,y)$ (Fig.2(b)) shows parallel strips along potential minima as per the periodicity of the external potential. This further confirms preferred particle positions at the modulation minima. $g_{ss}(x,y)$ in Fig.2 (c) shows dark patches on overall circular pattern, signifying slight deviation from isotropic structure. Due to breaking of clusters, the smaller particles  arrange themselves nearer the bigger species. This leads to peaked structure in the big-small correlations $g_{bs}(x,y)$ (Fig. 2(d)), suggesting enhanced mixing tendency between the two species which are phase separated in the absence of the external potential.

The alignment tendency of the particles is quantified by the relative orientation ($\theta_{ij}$) between $i^{th}$ and $j^{th}$ particles with respect to modulation direction,  $\theta_{ij}=tan^{-1}(\frac{y_j-y_i}{x_j-x_i})$.  We show in Fig. 2(e) and (f) the distribution $P(\theta)$ of bond angle $\theta_{ij}$ considering all the  pairs over equilibrium configurations. Data for the big particles $P^{(b)}(\theta)$ and small particles $P^{(s)}(\theta)$ are shown in Fig. 2(e) and (f) respectively. Peaks in $P^{b}(\theta)$ are at $\theta = 90^{\circ}$ and $270 ^{\circ}$, indicating alignment along the minima of the external potential. The smaller peaks in $P^{b}(\theta)$ at $\theta = 0^{\circ}$, $180^{\circ}$ and $360^{\circ}$ are due to  neighbouring big particles. $P^{(s)}(\theta)$ versus $\theta$ for smaller particle (Fig. 2(e)) shows peaks at integral multiples of $30^{\circ}$. This suggests that the smaller particles are forming clusters with a local hexagonal order among themselves, similar observation that reported in earlier experimental work \cite{paper 37}. 

All the tendencies enhance with increasing  $ \beta V_0$ as can be seen in SI show stronger alignment of bigger particles along the potential minima.(Fig. SI 1(a)) The clusters of the small particles further break up into smaller clusters, thus the big-small species shows enhanced tendency of mixing together (Fig. SI 1(b)-(d)).  Even the smaller particles also tend to get aligned along the potential minima now. The alignment tendency is further reflected by the strong peak of in both $P^{b}(\theta)$(Fig. SI 1(e)) and $P^{s}(\theta)$(Fig. SI 1(f)) at $\theta=90^{\circ}$.

{\it Mixing behaviour}

We further quantify the mixing tendency in the presence of the external modulation. First, we characterize the breaking of the small particle clusters. We compute the size of clusters $c$ of the small particles as follows. We take the  small particles to belong to a cluster whose  centre to centre distances in x- and y-direction are less than $x_{cl}$ and $ y_{cl}$ respectively. We choose $x_{cl}=1.1$ and $y_{cl}=1.25$. Here $x_cl$ corresponds to the first minimum in $g_ss(r)$ and $y_cl$ is slightly larger. We show distribution $P(c)$ for smaller particles in Fig. 3(a)-(b) for two different $V_0$'s. Large clusters of the smaller particles are observed in agreement to phase separation between two species in absence of external potential. It is clear that the distribution peaks are shifted to lower values of cluster sizes $c$ for $\beta V_0(=10.0)$(Fig. 3(b)).  

We plot the mean cluster size $\langle c \rangle (=\langle c \rangle = \int cP(c)dc)$ versus $\beta V_0$ in Fig. 3(c). We observe that $\langle c \rangle =Ae^{-B(\beta V_0)}$ with $A=-0.81$ and $B=4.8$ as shown in the inset of Fig. 3(c). The exponential decay suggests that energy needed to break the clusters is supplied by the external modulation. Note that $<c>$ falls to $\frac{1}{e}$ at $V^*=\beta V_0=1/B$ as shown in Fig. 3(d). We observe that  $V^*\sim\epsilon_{bs}^{2.4}$ in the inset of fig.3(d).  

We  calculate the demixing order parameter for different $\beta V_0$. We partition to this end the simulation box into small rectangular cells with dimension $\Delta x=1.1$, $\Delta y=1.25$ as per the appearance of first peak in anisotropic PCF data. Let $n_{ib}$ and $n_{is}$ be the numbers of big and small particles in each of the boxes respectively. The demixing order parameter is defined as:
\begin{equation}
O_d=\frac{1}{N}\Sigma_{i=1}^{M}|(n_{ib}-n_{is})|.
\end{equation}
Here $N$ is the total number of particles and $M$ the total number of partitions of the system. Fig. 4(a) shows the distribution $P(O_d)$ for all three cases for different $\beta V_0$. We observe a prominent peak at large $O_d$ value for $\beta V_0=0.0$. The peak shifts to a lower $O_d$ value, indicating mixing enhanced with $\beta V_0$. We show in Fig. 4(b) the  order parameter averaged over the configurations $\langle O_d \rangle$ vs $\beta V_0$. We observe that $\langle O_d \rangle$ decreases with $\beta V_0$ following two distinct branches, one branch for low values and the other one for higher values with discontinuity around $\beta V_0$=3.0. 

We also simulate the system behaviour for different system sizes, namely, N=144 and N=4096. The data for $\langle O_d \rangle$ in Fig. 4(c) and 4(d) also show discontinuity. The discontinuity in $\langle O_d \rangle$, $\Delta$  is estimated as the gap between the points where the discontinuity starts at the upper branch and  the lower branch. Fig. 4(e) shows the gap $\Delta$ for different N. We observe that this discontinuity increases systematically, confirming a first order transition. 

We also examine the behaviour of the specific heat of the system,  using the fluctuation of energy: 
\begin{equation}
C_v=<\frac{(E-\bar{E})^2}{N}>,
\end{equation}
Here E is the energy per particle at a given configuration and $\bar{E}$ is the average energy per particle.  Fig. 4(f) shows the $C_v$ vs. $\beta V_0$ plot. We observe peak in specific heat at  $\beta V_0$=3.0 where  the discontinuity in $\langle O_d \rangle$ is located. 

The peak in specific heat data, shown in Fig.4(f), show shifts to larger $\beta V_0$ with increasing as $N$, although the point of the order parameter discontinuity does not seem to depend sensitively on N.  The  peak values, of the heat capacity  $C_{max}$ vs. $N$ plot  in inset of Fig. 4(f) show linear dependence on $N$ dependence, as in case of first order transition\cite{paper 39}. Thus, system undergoes first order phase transition while going from a complete phase separated state to a mixed one in presence of an external modulation potential.

Physically, the big particles tend to get aligned more strongly than the small ones, making their ways splitting through the clusters of smaller particles with increasing $\beta V_0$. The surface energy cost of this breaking up is supported by the external potential. The minimum value of this external potential above which the cluster will break can be estimated qualitatively.  Let us consider for simplicity the case that a cluster of radius R breaks into two of radii $r_1$ and $r_2$. The change in line energy is given by:$\Delta f = 2\pi \sigma(r_1+r_2-R)$ where $\sigma$ is the coefficient of line tension. The minimum of the change in line free energy can be estimated by minimizing with respect to one radius $r_1$ and $r_2$ if we  assume that the mean density $\rho$ of the particles do not change due to breaking cluster and no particle is lost during the break up. This implies that $r_1^2+r_2^2=R^2$. Using this relation and minimizing with respect to $r_1$, we find $\Delta f_{min}=2\pi\sigma R(\sqrt{2}-1)$. The droplet radius $R$ consisting of small particles  in the background of the large particles can be stabilized if the total free energy consisting of the bulk and line tension component is at minimum. If $\delta g$ is the free energy per unit area of the droplet, then the minimum free energy corresponds to $R=\sigma/\delta g$. Hence, $\Delta f_{min}\sim \sigma^{2}$. The line tension experienced by the droplet of the smaller particles in the background of the larger particles is proportional to $\epsilon{bs}$ and consequently, $\Delta f_{min}\sim\epsilon^2_{bs}$. This amount of energy needs to be supplied through the external potential, implying that $\beta V_{0}\sim\epsilon^2_{bs}$. This is consistent to the simulation observation  of $V^*\sim\epsilon_{bs}^{2.4}$ dependence.

Demixing phase transition is well known in binary systems. Typically demixing is known to show phase coexistence ending in a critical point\cite{paper 40}. The critical point for demixing of binary colloids tuning the size ratio has  been reported \cite{paper 41}. Shifting of critical point in case of a critical colloidal-liquid to colloidal-liquid demixing phase transition can be controlled by solvent temperature\cite{paper 42}. Experiment on suspensions of colloids of differing charge but similar size undergoing phase separation into crystal and fluid phases has also been reported previously\cite{paper 43}. Phase transition due to demixing between phases rich in high-charge and low-charge colloids is observed in this case. First order phase transition in demixing due to size anisometry in colloids has been reported as well\cite{paper 44}. It may be noted that in contrast to the earlier reports, we observe here mixing of demixed binary colloids via a first order phase transition induced by an external modulation .

\section{Conclusion}

To summarize, we study using the MC simulations the mixing behaviour of a binary  colloidal film periodically modulated in one direction. The interplay between interaction and external potential results in an enhanced tendency of mixing among the two species which are de-mixed in absence of the modulation. The changes from de-mixed to mixed  state is accompanied by a first order phase transition with a jump in de-mixing order parameter. Our results suggest that mixing in a binary colloidal system can be tuned by suitable external potential which may be relevant in  technological applications. It will be interesting to study this phase transition including the critical behaviour, if at all, in future.

\section{Acknowledgements}

 The authors declare no conflict of interest. S. P. thanks computational centre at SNBNCBS for providing with the high computing facility and DST for financial support through the S N Bose PhD program.

\newpage

\begin{figure}[H]
	\centering
	\includegraphics[width=10cm]{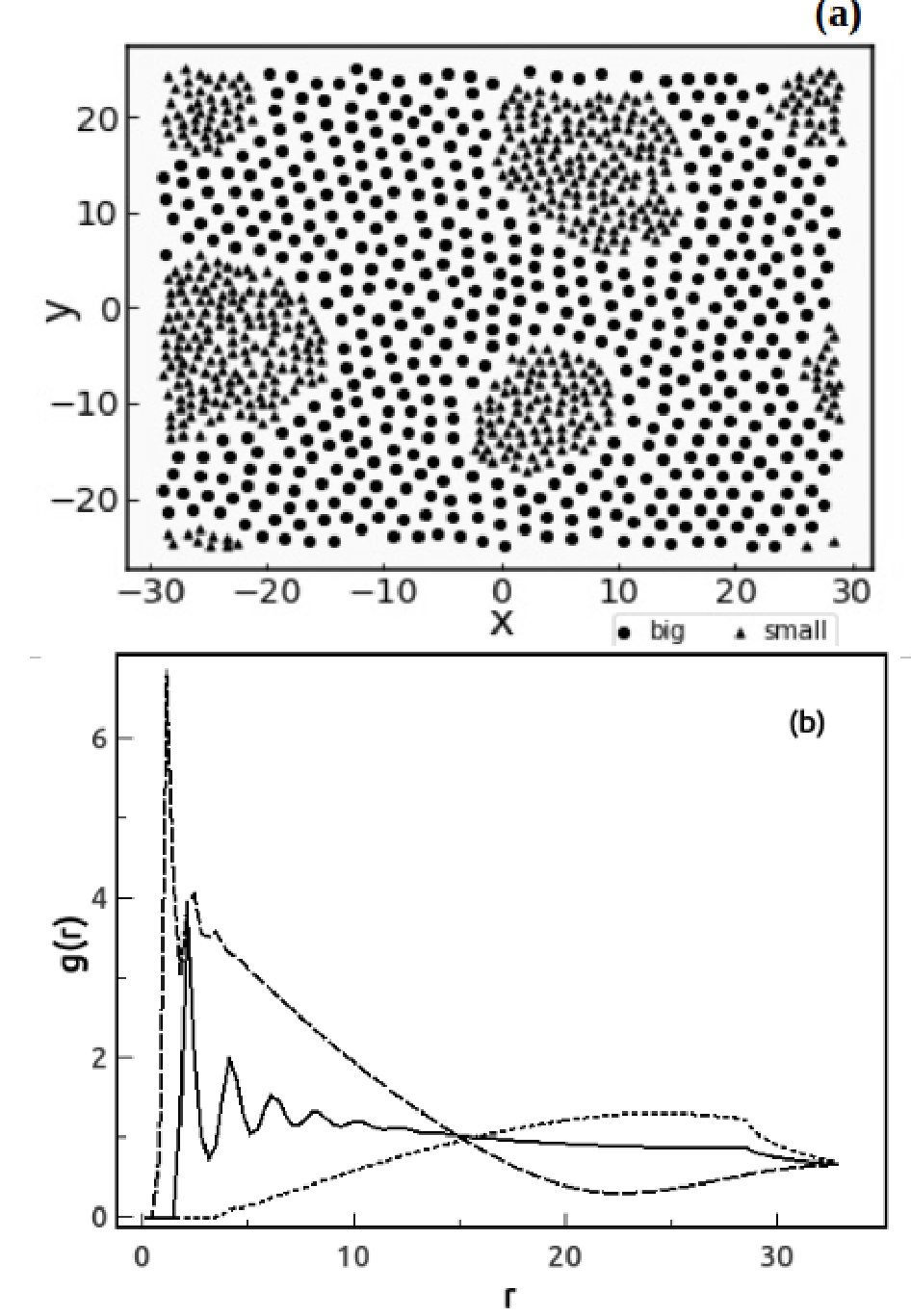}
    \caption{(a) final configuration at $\beta V_0=0.0$, Isotropic PCF at $\beta V_0=0.0$ for big-big ,  small-small, big-small interaction.}
\end{figure}

\begin{figure}[H]
	\centering
	\includegraphics[width=16cm]{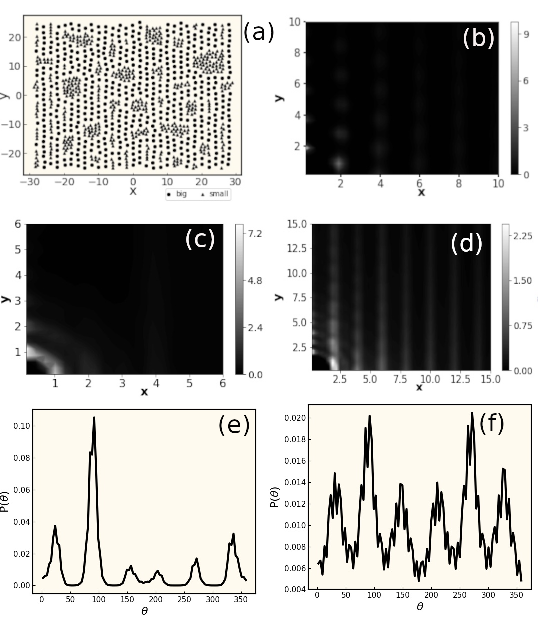}
    \caption{(a) final configuration at $\beta V_0=5.0$, anisotropic PCF at $\beta V_0=5.0$ for (b) big-big , (c) small-small, (d) big-small interaction. (e) Big-big bond angle distribution and (f) small-small bond angle distribution for $\beta V_0=5.0$.}
\end{figure}

\begin{figure}[H]
	\centering
	\includegraphics[width=20cm]{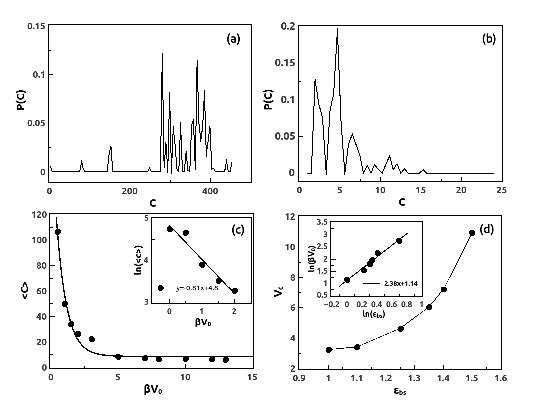}
     \caption{(a) Anisotropic cluster size distribution of smaller particles at $\beta V_0=0.0$, (b) $\beta V_0=10.0$. (c) Average cluster size vs. $\beta V_0$, with inset semi log fitting of $\langle c\rangle$ vs. $\beta V_0$. (d) $\beta V_0$ vs. $\epsilon _{bs}$ with inset log-log fitting of the same.}
\end{figure}

\begin{figure}[H]
	\centering
	\includegraphics[width=16cm]{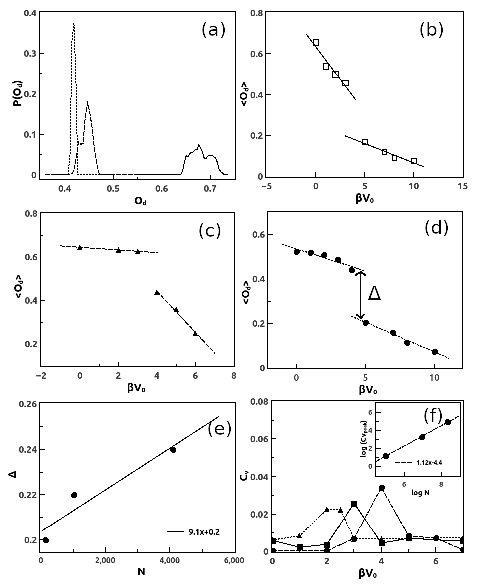}
    \caption{(a) Distribution of demixing order parameter ($O_d$) at$\beta V_0=0.0$ (solid), $\beta V_0=5.0$(dashed), $\beta V_0=10.0$(dotted line). (b) Average order parameter ($<O_d>$) vs. $\beta V_0$ for N=1024(solid,square), (c) N=144(dotted,triangle) and (c) N=4096 (dashed line,circle). (e) Gap $\Delta$ vs. N , (f) Specific heat $C_v$ vs. $\beta V_0$ for N=144(triangle, dotted), N=1024(square, solid), N=4096(circle, dashed line) with log-log fitting of heat capacity $Cv_{peak}$ vs. N in the inset.}
\end{figure}


\newpage

\section{Supplementary data}

The snapshot showing enhanced mixing of particles under higher modulation is shown in Fig. S1(a). The bigger particles have been strongly aligned along the potential minima while breaking the clusters of the smaller particles into further smaller one; hence an enhanced tendency of mixing is observed. The anisotropic PCF data taking into account for three different interaction is shown in Fig S1(b)=(d). $g_{bb}(x,y)$ (Fig.2(b)) shows strong allignments of paticles through the parallel strips along potential minima as per the periodicity of the external potential. This confirms strong preference of particle positions at the modulation minima. $g_{ss}(x,y)$ in Fig. S1(c) shows distorted patches on overall circular pattern, signifying more deviation from isotropic structure. Due to the mixing phenomena, the smaller particles arrange themselves nearer the bigger species more so now. So the big-small correlation shows $g_{bs}(x,y)$ (Fig. S1(d)) higher peak values, suggesting enhanced mixing tendency between the two species with increment of strength of the external potential.

The alignment tendency is further reflected by the strong peak of in both $P^{b}(\theta)$(Fig. SI 1(e)) and $P^{s}(\theta)$(Fig. SI 1(f)) at $\theta=90^{\circ}$ big and small particles' arrangement with respect to the modulation direction. Two significant peaks dominate in both the cases signifying strong alignment perpendicular to the modulation direction, i.e. along the potential minima.

\begin{figure}[H]
	\centering
	\includegraphics[width=14cm]{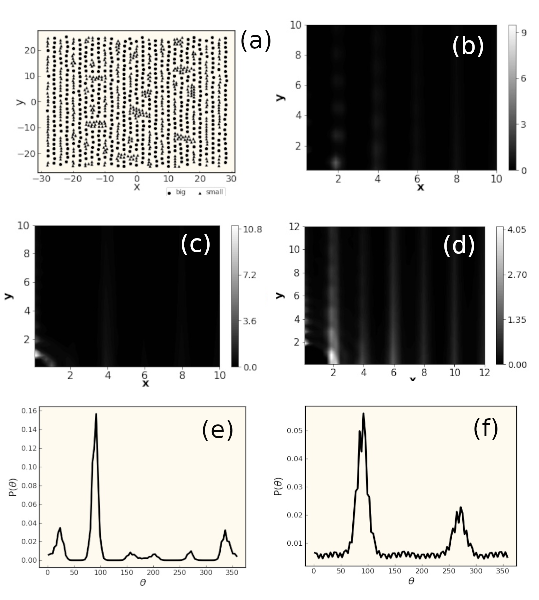}
    \caption{(a) final configuration at $\beta V_0=10.0$, anisotropic PCF at $\beta V_0=10.0$ for (b) big-big , (c) small-small, (d) big-small interaction. (e) Big-big bond angle distribution and (f) small-small bond angle distribution for $\beta V_0=10.0$.}
\end{figure}

\end{document}